\documentclass[12pt, a4paper]{article}
\usepackage[margin=1in]{geometry}
\usepackage[T1]{fontenc}
\usepackage{tgtermes}
\usepackage{tgheros}
\usepackage{tgcursor}
\usepackage{amsmath, amsthm, amssymb}
\usepackage{mathtools}
\mathtoolsset{showonlyrefs=true}

\usepackage{booktabs,array}
\usepackage{xcolor}
\usepackage{graphicx}
\usepackage{float,placeins}
\usepackage{xurl}
\usepackage[
    colorlinks=true,
    citecolor=blue!50!black,
    urlcolor=blue!50!black,
    linkcolor=blue!50!black,
    pdftitle={Feedback Design in the Shadow of Strategic Communication},
    pdfsubject={Self-contained derivations and prior-dependent disclosure policies}
]{hyperref}
\usepackage{secdot}
\usepackage{enumitem}
\setlist[enumerate]{itemsep=0.25em,topsep=0.4em}

\newtheorem{proposition}{Proposition}
\newtheorem{lemma}{Lemma}
\newtheorem{corollary}{Corollary}
\theoremstyle{definition}

\newtheorem{assumption}{Assumption}
\theoremstyle{remark}

\newcommand{\E}{\mathbb E}
\newcommand{\cav}{\operatorname{cav}}

\newcommand{\var}{\operatorname{Var}}

\usepackage{setspace}

\usepackage{natbib}
\title{\bf Feedback Design in the Shadow of Strategic Communication\footnote{This study was financially supported by the JSPS KAKENHI Grant Number 26K16320.}}
\author{Daiki Kishishita\thanks{Graduate School of Economics, Hitotsubashi University. E-mail: \url{daiki.kishishita@gmail.com}} \qquad Tomoki Tsujita\thanks{Hitotsubashi Institute for Advanced Study, Hitotsubashi University. E-mail: \url{tomoki.tsujita@r.hit-u.ac.jp}}}
\date{}

\begin{document}
\maketitle

\begin{abstract}
    We study optimal feedback design when evaluations affect subsequent strategic communication within organizations. A firm designs public feedback about an employee's potential before the employee communicates private information about project-specific circumstances to a manager. While more informative feedback helps the manager make better decisions, it can also intensify preference conflicts and undermine the credibility of subsequent communication. We characterize the optimal feedback policy. First, full disclosure is either optimal for all prior beliefs or strictly suboptimal for all of them. In particular, it is optimal when preference conflicts are sufficiently small or sufficiently large. Second, when full disclosure is suboptimal, no disclosure is optimal for some priors, while partial disclosure is strictly optimal for some priors if and only if the number of messages in equilibrium communication varies with posterior beliefs. Third, optimal partial disclosure can conclusively identify low-potential employees when high potential is rare, but high-potential employees when it is common.
    \begin{description}
        \item[Keywords:] Information design; Cheap talk; Feedback; Employee assessments
        \item[JEL Classification:] D82, D83, D23.
    \end{description}
\end{abstract}

\newpage
\begin{spacing}{1.2}

\section{Introduction}
Determining how ambitious the projects assigned to employees should be is a fundamental challenge in organizational management. The appropriate level of ambition depends critically on employees' potential, which may not be fully known to either employees themselves or their managers.  To reduce this
uncertainty, many organizations use formal assessments of employees'
capabilities, and share the results with employees and their managers. 

One prominent example is the
\emph{assessment center}, which evaluates employees through structured
exercises, such as work simulations and interviews \citep{dimotakis2017positive}. For instance,
PepsiCo's Leadership Assessment and Development program uses
a multidimensional assessment process to evaluate 
potential.\footnote{\url{https://www.td.org/content/td-magazine/pepsicos-formula-for-leadership-potential}}
In a survey of 80 organizations selected for their talent-development
practices, \cite{church2015top} report that 80\% used assessments
of high-potential employees or senior executives.\footnote{Furthermore, \cite{church2015top} report that 17\%, 39\%, and 33\% of respondents made complete assessment results, integrated summaries, and topline summaries, respectively, available to both employees and their managers (multiple responses were allowed). This indicates that sharing assessment information with both employees and their managers is an established practice among  organizations.} These practices
raise a fundamental question: Should firms disclose assessments of
employee potential to employees and their managers, and, if so,
how should such feedback be designed?

The answer to this question depends not only on what assessments reveal about employees' potential, but also on how such feedback affects the communication of other information relevant to assignment decisions. This other information concerns project-specific circumstances, which are often better known to employees directly involved in the relevant activities than to the managers who make final decisions. For example, a salesperson may have better information about customer demand, which affects how ambitious a sales target should be, while an R\&D employee may better understand technical constraints that affect the appropriate scope of a project. Managers must therefore rely on employees
to communicate what they know about particular opportunities and constraints. How effectively such dispersed information can be communicated to decision makers is a central concern in organizational economics \citep{dessein2002authority, alonso2008does}.

Yet such communication need not be fully informative, as employees may prefer more ambitious projects than would be optimal for their firms \citep{dessein2002authority}. In particular, challenging projects can provide opportunities
to develop skills that remain valuable beyond the current employment
relationship. Employees may therefore value ambitious assignments
more highly than their managers do. When local circumstances are
not directly verifiable, this conflict of interest can lead employees
to exaggerate favorable information in an attempt to obtain more
challenging assignments.

Crucially, the extent of this conflict depends on feedback about
employee potential. An unfavorable assessment may reduce an employee's
expected benefit from challenging work, narrowing the disagreement
with the manager. A favorable assessment, by contrast, may strengthen
the employee's desire for such assignments and widen the disagreement.
Feedback therefore plays a dual role. It directly improves managerial decisions by reducing uncertainty about employees' potential, but it also indirectly affects those decisions by changing the credibility of employees' subsequent communication. More informative feedback may consequently lead to less informed decisions.

Although existing theoretical studies of feedback have
examined its effects on employees' effort and other direct choices
\citep[e.g.,][]{goltsman2011interim, ely2025feedback},
they have paid less attention to its effects on subsequent strategic
communication.\footnote{Empirical studies also focus on its effect on employees' learning, effort, and psychological
reactions (see \cite{anseel202525} for a review).} This paper develops a theoretical framework to study
how firms should design feedback when both effects matter.

For this purpose, we combine a standard cheap-talk model
in a uniform-quadratic environment \citep{crawford1982strategic}
with Bayesian persuasion \citep{kamenica2011bayesian}. Our model
features a designer (a firm), a sender (an employee), and a
receiver (a manager). The sender privately observes project-specific
circumstances and communicates this information to the receiver
through cheap talk. The receiver then chooses an action, interpreted
as the employee's project assignment. Both players' ideal
actions depend on the project-specific circumstances and the employee's
potential, but the latter receives a greater weight in the sender's
preferences than in the receiver's. Consequently, the sender has
an incentive to exaggerate favorable information to obtain a more
ambitious assignment. 

Employee potential is initially unknown to
both players. Before communication takes place, the designer commits
to an information structure that generates a public signal about
the employee's potential. The designer's problem is to choose this
feedback policy to minimize the firm's expected decision error,
taking into account its effect on the subsequent communication game.\footnote{As in standard cheap-talk games, the continuation communication game has multiple equilibria. We assume that the most informative equilibrium is played both on and off the equilibrium path. This selection is Pareto efficient for all players.}  Crucially, feedback design can affect the informativeness of subsequent communication by changing the number of messages in the most informative equilibrium. Thus, feedback not only provides information about employee potential but also shapes how much project-specific information the manager can learn from the employee.

Our analysis yields three main results. First, we establish a
necessary and sufficient condition for full disclosure to be
optimal. Its optimality is independent of the prior distribution of employees' potential: full disclosure is either optimal
for every interior prior or strictly suboptimal for all of them.
We further show that full disclosure is optimal when the
sender--receiver preference conflict is sufficiently small
and also when it is sufficiently large. When the conflict is
small, communication remains sufficiently informative that
revealing employee potential is beneficial. When the conflict
is sufficiently large, communication is already uninformative,
so revealing potential cannot worsen it further. At
intermediate levels of conflict, however, withholding some
information can improve decision-making.

Second, when full disclosure is suboptimal, no disclosure is
optimal for a nonempty interval of prior beliefs. Moreover,
partial disclosure is strictly optimal for some priors if and
only if the number of partitions in the continuation
communication equilibrium varies with the posterior belief. Furthermore, 
we characterize how the optimal policy changes with the prior distribution of employees' potential. In particular, there are at
most two separate intervals of priors for which no disclosure
is optimal. Consequently, the optimal policy has at most five
successive regions, alternating between partial disclosure
and no disclosure. Thus, the optimal degree of feedback need not vary monotonically with the probability of an employee having high potential.

Third, we characterize the information conveyed by optimal
partial disclosure. When the prior probability of high
potential is relatively low, an optimal experiment can
conclusively identify low-potential employees while leaving
favorable assessments inconclusive. When high-potential
employees are relatively common, the opposite pattern can
emerge: favorable assessments conclusively identify high
potential, while unfavorable assessments remain inconclusive.
Furthermore, in a middle partial-disclosure region, neither
assessment need identify an employee's potential with
certainty. Thus, optimal partial disclosure can take
qualitatively different forms depending on the composition
of the workforce.

These results have several implications for the design of
employee evaluations. First, our results provide a novel rationale for withholding
assessments of employee potential. Such confidentiality
is observed in practice. For example, \citet{lee2026potential}
document two public-sector organizations in Singapore
where detailed assessments of employees' career potential
are largely withheld from frontline employees and their
managers, with access restricted to senior management
and HR. Our analysis suggests that such practices may
improve managerial decisions by preserving the
informativeness of subsequent strategic communication.

Second, the
optimal evaluation standard may be relatively \emph{lenient}
when high-potential employees are rare, but more
\emph{stringent} when they are common. Under a lenient
standard, unfavorable evaluations are decisive, whereas
favorable evaluations are less discriminating. Under a
stringent standard, favorable evaluations are conclusive
but more difficult to obtain. 

Finally,
limiting feedback is particularly relevant when the
employee--manager preference conflict is intermediate:
neither so small that communication is already highly
informative nor so large that meaningful communication
has broken down.

Beyond these managerial implications, our analysis overcomes two technical difficulties.  First, we study information
design in an environment where the designer does not seek
to induce a particular action by the receiver. Instead,
the designer seeks to improve decision-making and may
have preferences perfectly aligned with those of the
receiver. Information disclosure nevertheless has a
strategic effect because it changes the incentives
governing subsequent cheap-talk communication.
Second, we provide a global characterization of optimal
information design despite the complexity introduced by
endogenous changes in communication equilibria. As the
posterior belief varies, the number of equilibrium
partitions changes discretely, generating a payoff function
that is the upper envelope of multiple quadratic functions
and exhibits upward kinks. Although this function can have
multiple local extrema, we show that its concave envelope
has a restricted structure: there can be at
most two separate no-disclosure intervals.
This allows us to characterize globally optimal feedback
policies without restricting the class of information
structures available to the designer.

The remainder of the paper is organized as follows. We first review the related literature. Section 2 presents the model, and Section 3 characterizes the equilibrium of the continuation communication game. Section 4 examines the benchmark case in which the equilibrium number of messages is constant across posterior beliefs, while Section 5 characterizes the optimal feedback design in the general case. Section 6 concludes.

\paragraph{Related literature:} Our paper contributes to two strands of literature: feedback design and strategic communication.

First, our paper relates to the theoretical literature on feedback design and performance evaluations. A body of research examines how feedback affects employees' incentives and behavior, including effort provision, learning, and innovation \citep[e.g.,][]{goltsman2011interim,hansen2013performance,zabojnik2014subjective,ely2025feedback,hayashida2026transparency}. This literature shows that partial disclosure and lenient evaluations can be optimal for incentive reasons. Our paper identifies a different rationale for limiting feedback: its effect on subsequent strategic communication. In particular, we show that full disclosure is not always optimal, and in such a case, the optimal evaluation standard can shift from lenient to stringent as the prevalence of high-potential employees increases, because feedback changes the credibility of employees' subsequent communication rather than their effort incentives.

In this literature, our paper is closely related to \citet{ball2025scoring}, who shows that aggregating an agent's observable characteristics into a score can improve decision accuracy relative to full disclosure. In his model, restricting information discourages agents from manipulating the characteristics used for evaluation. By contrast, in our model, employees cannot manipulate the attribute being evaluated. Instead, feedback changes the informativeness of subsequent cheap-talk communication about a different variable. Thus, our paper identifies a distinct rationale for limiting information and characterizes how the optimal form of feedback depends on the distribution of employee attributes.

Second, our paper contributes to the literature on cheap talk, building on \citet{crawford1982strategic}. One strand of this literature examines how conflicts of interest arise endogenously before communication takes place \citep[e.g.,][]{alonso2008does,antic2020cheap}. Our paper instead examines how public information about an uncertain attribute changes the conflict of interest between the sender and the receiver. Another strand examines the effects of public information on cheap-talk communication. In particular, \citet{chen2012value} shows that more informative public signals about the state on which the sender reports can reduce the informativeness of subsequent communication.\footnote{See also \cite{ishida2016cheap} for the effect of the receiver's information on communication.} In contrast, we study public information about a distinct attribute of the sender, which changes the sender--receiver preference conflict and thereby affects communication about the underlying state. We further characterize how such public information should be optimally designed.

Our paper also relates to the literature on information design with cheap talk \citep[e.g.,][]{ivanov2010informational,krahmer2021information,ivanov2022cheap,kreutzkamp2025persuasion,lyu2026information}. These studies examine how the information privately acquired by a sender should be designed, taking into account the sender's incentives to communicate it credibly. In contrast, we study the optimal disclosure of  public information about a sender's attribute. Because the feedback is public, it changes their common beliefs and hence the preference conflict governing subsequent communication about a different variable. Thus, while the former studies design the sender's information acquisition to facilitate credible communication, our paper shows how public information disclosure can alter communication incentives and characterizes the resulting optimal disclosure policy.

The studies closely related to ours include \citet{jain2018bayesian} and \citet{lipnowski2022persuasion}. In \citet{lipnowski2022persuasion}, the sender publicly announces an experiment. With a fixed probability, the receiver observes a report generated by the experiment. With the remaining probability, the sender observes the state and chooses the report herself. The two channels thus convey information about the same underlying state, and the receiver is unsure of whether the report is generated by the experiment. In our model, the designer commits to public feedback about the employee's potential, and the sender subsequently communicates private information about project-specific
circumstances through cheap talk.

\cite{jain2018bayesian} characterizes optimal grading policies in a model with multidimensional binary states and a binary hiring decision. In her model, informative cheap talk is possible only for intermediate posterior beliefs, whereas its informativeness weakly decreases with the posterior mean in ours. Moreover, her designer is the sender, while ours may have the same objectives as the receiver. These differences yield distinct predictions: unlike in her  model, full disclosure and no disclosure can be uniquely optimal over a nondegenerate interval of priors. This allows us to identify when partial disclosure strictly outperforms both.

\section{Model}
We develop a model of information design with cheap-talk communication. The model consists of three players: a sender $S$, a receiver $R$, and a designer $0$. In our primary application, the sender is an employee, the receiver is a manager, and the designer is a firm. Although our model is not restricted to this setting, we use this application to motivate and interpret its key assumptions throughout the analysis. In the model, after the designer designs the feedback system as information design, the sender and the receiver play a cheap-talk communication game.

\paragraph{Communication game:} We begin with the model of the continuation communication game. At the beginning of the communication stage, the sender observes an underlying state $\omega$ of nature and sends a cheap-talk message to the receiver. After observing the message, the receiver chooses an action $y\in\mathbb R$, which represents the level of ambition of the project assigned to the employee. A higher $y$ corresponds to a more ambitious project.

The optimal choice of $y$ depends on two variables, $\omega$ and $\theta$. The former captures project-specific circumstances, while the latter represents an attribute of the sender, such as the employee's potential. We assume that $\omega$ is uniformly distributed on $[0,1]$. The sender observes $\omega$, whereas the receiver does not. Thus, $\omega$ is private information for the sender. This captures the idea that employees, who work directly on relevant projects, often possess better information about project-specific circumstances than their managers.

In contrast, $\theta\in\{L,H\}$, where $0<L<H$, represents an attribute that is initially unknown to both the sender and the receiver. The common prior probability that $\theta=H$ is $p\in(0,1)$. We assume that $\omega$ and $\theta$ are independently distributed.

The players' utilities are given by
\begin{equation}
    u_i ( y , \omega , \theta ) = - ( y - \omega - \alpha_i \theta )^2 ,
\end{equation}
for $i\in\{0,S,R\}$.
We normalize $\alpha_0=1$ without loss of generality.
Thus, differences in $\alpha_i$ capture the players' conflicts of interest. Throughout the paper, we assume $\alpha_S>\alpha_R>0$, so that the sender prefers a more aggressive choice of $y$ than the receiver does.
We also assume $\alpha_R<2$.
Because a preference difference appears repeatedly, we abbreviate it as
\[d\coloneqq\alpha_S-\alpha_R>0.\]
Importantly, we do not require a conflict of interest between the receiver and the designer. The results arise even when their preferences are perfectly aligned (i.e., $\alpha_R=1)$.

To illustrate, suppose that $y$ represents how demanding a project is and $\theta$ captures the employee's potential to develop human capital through challenging assignments. A higher $\theta$ makes a more demanding assignment desirable. Both the employee and the manager value human-capital development, but the employee values it more because the skills acquired remain useful even if she leaves the firm \citep{acemoglu1997training}. This creates a conflict of interest over project assignments.\footnote{Employees may also favor more ambitious projects because of career concerns. Successfully completing challenging assignments may allow employees to reveal their ability to future employers, who do not observe the firm's internal assessments of employee potential. Since higher-potential employees are more likely to succeed in such assignments, the reputational benefits of undertaking ambitious projects may increase with employee potential. This provides an alternative rationale for the preference divergence in our model and is closely related to \cite{ke2023learning}.}

\paragraph{Feedback design:}
Before the communication game, the designer commits to a feedback system that provides information about $\theta$ to both the sender and the receiver. Formally, the designer chooses a finite public feedback experiment
\[
\sigma:\{L,H\}\to\Delta({\cal S}),
\]
where ${\cal S}$ is a finite set of feedback signals and $\Delta({\cal S})$ denotes the set of probability distributions over ${\cal S}$. Conditional on $\theta$, a feedback signal $s$ is drawn according to $\sigma(\cdot | \theta)$ and publicly observed by both the sender and the receiver before the communication game begins. The sender and the receiver then update their beliefs about $\theta$ according to Bayes' rule. 

Let $q = \Pr ( \theta = H | s )$ denote the posterior probability that $\theta = H$ given $s$. The posterior mean of $\theta$ is given by
\begin{equation}
    x \coloneqq \E [ \theta | s ] = ( 1 - q ) L + q H ,
\end{equation}
whereas its prior mean is given by $\mu \coloneqq ( 1 - p ) L + p H$.

We distinguish three types of feedback experiments. \emph{No disclosure} leaves the posterior equal to the prior, $q=p$, with probability one. \emph{Full disclosure} reveals $\theta$ perfectly, inducing posteriors $q=0$ and $q=1$ with probabilities $1-p$ and $p$, respectively. Any informative experiment that does not fully reveal $\theta$ is called \emph{partial disclosure}. 

\paragraph{Timing of the game:} The timing is summarized as follows.
\begin{enumerate}
    \item The designer commits to a finite public feedback experiment $\sigma$.
    \item Nature draws $(\omega,\theta)$, and the public feedback $s$ is realized according to $\sigma ( \cdot | \theta )$.
    \item The sender observes $\omega$ and sends a costless message to the receiver.
    \item The receiver observes the message and chooses $y$.
\end{enumerate}

\paragraph{Equilibrium concept:}
For any feedback experiment $\sigma$ and any feedback signal $s$ that occurs with positive probability, we study a perfect Bayesian equilibrium (PBE) of the continuation communication game between the sender and the receiver. A PBE consists of the sender's messaging strategy, the receiver's action strategy, and the receiver's beliefs about $\omega$, such that both players' strategies satisfy sequential rationality and beliefs are updated according to Bayes' rule whenever applicable.

The continuation communication game generally admits multiple equilibria, including a babbling equilibrium \citep{crawford1982strategic}. As in the standard uniform-quadratic model, every equilibrium is outcome-equivalent to an \emph{interval-partition equilibrium} $( \{ \omega_i \}_{i=0}^n, \{ y_i \}_{i=1}^n)$, where $n$ is positive integer, $\{ \omega_i \}_{i=0}^n$ is a partition of $[ 0, 1 ]$ with $0=\omega_0<\omega_1<\cdots<\omega_n=1$, and each interval $[\omega_{i-1},\omega_i]$ induces a distinct action $y_i$.

To resolve equilibrium multiplicity, we impose the following equilibrium-selection assumption.

\begin{assumption}[Equilibrium selection]\label{ass:selection}
    For any posterior belief about $\theta$ induced by the feedback experiment, the players play the interval-partition equilibrium with the largest number of positive-probability messages.
\end{assumption}

The equilibrium with the largest number of positive-probability messages maximizes the expected payoffs of all three players and thus Pareto-dominates the other equilibria. Our assumption therefore selects the Pareto-efficient continuation equilibrium after each feedback signal.

Taking the selected continuation equilibria as given, the designer chooses a feedback experiment $\sigma$ to maximize her ex-ante expected payoff.

\FloatBarrier
\section{Equilibrium of the Communication Game}

We first characterize the equilibrium of the communication game following a feedback signal~$s$.
After this signal, the posterior mean of $\theta$ is given by $x$.

Conditional on $s$, the variance of $\theta$ does not affect either player's optimal choice. For any given $\omega$, the receiver's ideal action is $\omega+\alpha_Rx$, whereas the sender's is $\omega+\alpha_Sx$. Their difference is $dx$, independent of $\omega$. Thus, a higher posterior mean of $\theta$ increases the conflict of interest between the sender and the receiver, potentially making communication less informative.

\subsection{Equilibrium Characterization}
Conditional on $s$, our communication game is equivalent to the quadratic-uniform cheap-talk model of \citet{crawford1982strategic}, with preference bias $dx$. The following lemma characterizes interval-partition equilibria.

\begin{lemma}[Crawford and Sobel: Cheap-talk equilibrium]
\label{lem:capacity}
For any posterior mean $x$, an equilibrium with $n\ge2$ distinct induced actions exists if and only if
\begin{equation}\label{eq:feasible}
2dx\,n(n-1)<1. \notag
\end{equation}
In such an equilibrium, the state space is partitioned into $n$ intervals with boundaries
\begin{equation}\label{eq:cutoffs}
\omega_i=\frac{i}{n}+2dx\,i(i-n),
\qquad i=0,\ldots,n. 
\end{equation}
The length of interval $i$ is
\begin{equation}\label{eq:lengths}
\ell_i\coloneqq\omega_i-\omega_{i-1}
=\frac1n+2dx(2i-n-1),
\qquad i=1,\ldots,n. 
\end{equation}
Upon receiving the message corresponding to interval $i$, the receiver chooses
\begin{equation}\label{eq:receiver}
y_i=\frac{\omega_{i-1}+\omega_i}{2}+\alpha_Rx. 
\end{equation}
\end{lemma}

Under Assumption~\ref{ass:selection}, the equilibrium number of intervals (or messages) is therefore
\begin{equation}\label{eq:N}
    N ( x ) \coloneqq \max \{ n \ge 1 \colon 2 d x \ n ( n - 1 ) < 1 \} .
\end{equation}
Hence, $N(x)$ is weakly decreasing in $x$: a higher posterior mean increases preference conflict and reduces the amount of information that can be credibly communicated. This is the key property behind our results.

For $n\ge2$, define
\begin{equation}
    t_n\coloneqq\frac1{2dn(n-1)} \in ( L , H ).
    \label{eq: threshold of posterior mean}
\end{equation}
$t_n$ is a threshold of the posterior mean at which the number of intervals changes from $n$ to $n - 1$
Then, $N(x) = n$ for $x\in[t_{n+1},t_n)$, while $N(x)=1$ for $x\ge t_2$. At each threshold $t_n$, the equilibrium number of distinct induced actions falls from $n$ to $n-1$.

\subsection{The Designer's problem}


We now derive the designer's expected payoff following a feedback signal.
With $n$ equilibrium messages, the receiver's action reflects both the midpoint of the reported interval and the posterior mean $x$. Accordingly, the designer's loss consists of two components: the adaptation loss with respect to $\omega$ and that with respect to $\theta$. Since the within-interval error in $\omega$ has conditional mean zero, the cross term between the two adaptation errors vanishes in expectation.
Thus, the designer's conditional expected payoff given $s$ can be written as 
\[\E [ u_0 | s ] = - \underbrace{\E \left[ \left. \left( ( \omega_i + \omega_{ i - 1 } ) / 2 - \omega \right)^2 \right| s \right]}_{\rm communication \ loss} - \underbrace{\E[(\alpha_Rx-\theta)^2| s]}_{\rm attribute\text{-}related \ decision \ loss}.\]

First, the expected loss from imperfect communication about $\omega$ can be rewritten as
\begin{equation}
    \E \left[ \left. \left( \frac{ \omega_i + \omega_{ i - 1 } }{ 2 } - \omega \right)^2 \right| s \right]
    = \frac1{12n^2}+\frac{ ( dx )^2(n^2-1)}{ 3 }
    =: \mathcal L_n ( x ).
    \label{eq: communication loss}
\end{equation}
The first term captures the loss that arises even in the absence of preference conflict, which decreases with the number of messages.
The second term captures the additional loss arising from the conflict of interest between the sender and the receiver.
For a given number of messages $n \geq 2$, this loss increases with $x$, as a higher posterior mean widens the preference conflict.

In contrast, the decision loss associated with the attribute $\theta$ is
\begin{equation}
\begin{aligned}
    \E[(\alpha_R x-\theta)^2 | s]
    &= {\rm Var}(\theta | s) + (\alpha_R-1)^2 x^2 \\
    &= \E[\theta^2 | s]-\alpha_R(2-\alpha_R)x^2.
\end{aligned}
    \label{eq: attribute loss}
\end{equation}
The first term captures the loss arising from residual uncertainty about $\theta$, while the second term captures the loss arising from the preference difference between the designer and the receiver.
In particular, when $\alpha_R=1$, the preferences of the designer and the receiver are aligned, and the loss reduces to
${\rm Var}(\theta | s)$, which is zero under full disclosure.
When $\alpha_R \neq 1$, by contrast, some loss remains even under full disclosure due to the preference difference.
Although the loss need not be monotonic in $x$, more informative feedback reduces its ex-ante expectation since $0<\alpha_R<2$.%
\footnote{
    To see this, observe that $\E_s[(\alpha_R x-\theta)^2] = \E[\theta^2]-\alpha_R(2-\alpha_R)\E_s[x^2]$.
}

Combining the two losses, the designer's conditional expected payoff given $s$ is
\begin{equation}
    \E[ u_0 | s ] = W_n ( x ) - \E [ \theta^2 | s ] ,
    \label{eq: V}
\end{equation}
where
\begin{equation}
    W_n(x)\coloneqq-\frac1{12n^2} + k_nx^2  \label{eq_wn}
\end{equation}
with
\begin{equation}
    k_n \coloneqq \alpha_R(2-\alpha_R) - \frac{d^2(n^2-1)}3 . 
\end{equation}
For any feedback experiment, the law of iterated expectations implies $\E_s \left[ \E[ \theta^2 | s] \right] = \E[ \theta^2 ]$.
Thus, the ex-ante expectation of the second term in \eqref{eq: V} is independent of the feedback experiment.
Hence, for a given number of messages $n$, the designer can evaluate feedback experiments based solely on $W_n(x)$.

The continuation communication game generally admits multiple equilibria for a given posterior mean $x$. Under Assumption \ref{ass:selection}, we focus on the equilibrium with the largest number of messages. Accordingly, the design-relevant payoff is given by $W ( x ) \equiv W_{ N ( x ) }$, where $N(x)$ denotes the maximum number of equilibrium messages for a given $x$.

In the feedback design problem, however, the designer chooses an experiment that changes the distribution of $x$, thereby affecting the number of equilibrium messages in the continuation game. This dependence on $N(x)$ appears to complicate the designer's problem. In particular, because the number of equilibrium messages changes discretely with $x$, the function $W$ may appear to have kinks at the thresholds $t_n$. The following lemma shows, however, that these kinks satisfy a regularity that will be useful in characterizing the optimal feedback policy.

\begin{figure}[t]
\centering
\includegraphics[width=.85\linewidth]{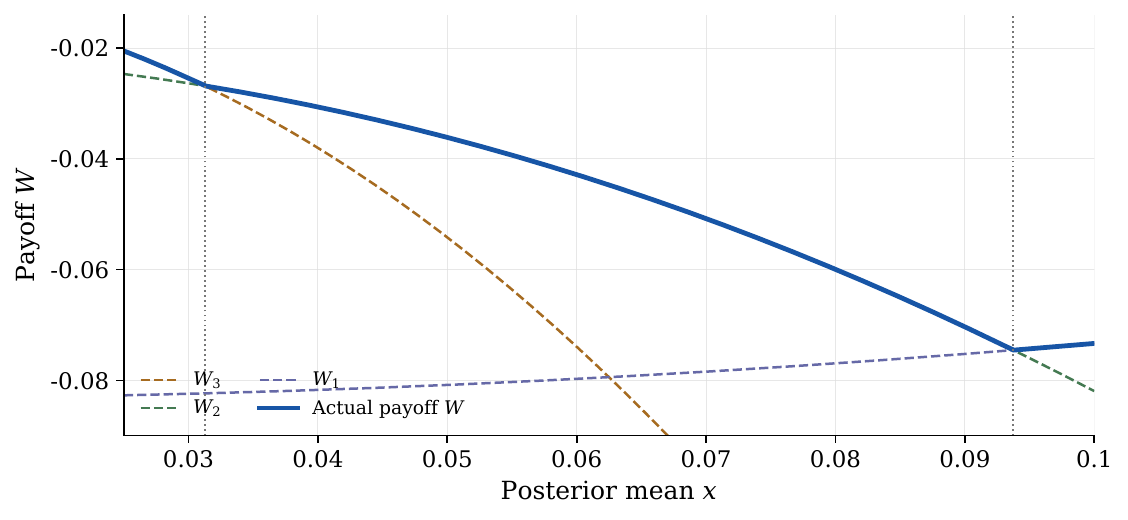}
\caption{Designer’s payoff. $\alpha_R=1$, $\alpha_S=11/3$, $L=1/40$, and $H=1/10$.}
\label{fig:pieces}
\end{figure}

\begin{lemma}[Regularity of $W$]
    \label{lem:envelope}
    For any $x \in [L,H]$,
    \begin{equation}
        W(x)=W_{N(x)}(x)=\max_{1\le n\le N(L)}W_n(x).
    \label{eq:maxrepresentation}
    \end{equation}
    The function $W$ is continuous. At every threshold $t_n\in(L,H)$
    where the equilibrium message count falls from $n$ to $n-1$, the slope of $W$ increases:
    \begin{equation}
        W_{ n - 1 }' ( t_n ) - W_n' ( t_n )
        = \frac{ d ( 2 n - 1 ) }{ 3 n ( n - 1 ) } > 0.
    \label{eq:kink}
    \end{equation}
\end{lemma}

The lemma has two important implications.
First, \eqref{eq:maxrepresentation} provides a convenient representation of $W$ as the upper envelope of the functions $W_n$, even when these functions are evaluated outside the regions where the corresponding $n$-message equilibria exist.
Note that the maximization in \eqref{eq:maxrepresentation} is a representation of the selected equilibrium payoff, rather than a choice over equilibria.
For example, Figure~\ref{fig:pieces} illustrates the case $N(L)=3$, where $W(x)=\max\{W_1(x),W_2(x),W_3(x)\}$.

Second, although $W$ is continuous, it exhibits an upward kink whenever the equilibrium number of messages decreases.
For instance, at the first vertical dotted line in Figure~\ref{fig:pieces}, the number of messages falls from three to two, and the relevant payoff switches from $W_3$ to $W_2$.
The lemma establishes that the slope of $W$ always increases at such thresholds.
This property will play a crucial role in characterizing the optimal feedback design.

We now formulate the designer's information design problem.
In our binary-state setting, the posterior mean $x=L+(H-L)q$ uniquely determines the posterior belief $q$. Moreover, the designer's conditional expected payoff depends on the feedback signal only through $x$. Thus, the designer's problem can equivalently be formulated as choosing a distribution of posterior means.
Let $x_j \in [ L , H ]$ denote the posterior mean of $\theta$ conditional on feedback signal $s_j$, and let $F_j \in [ 0 , 1 ]$ denote the probability of inducing $x_j$ with $\sum_{ j } F_j = 1$.
The designer's problem is
\begin{equation}
    V ( \mu ) = \max_{(x_j,F_j)_j}\sum_jF_j W(x_j)
    \quad \text{s.t.} \quad
    \sum_jF_jx_j=\mu .
\end{equation}
The constraint is the \emph{Bayes plausibility}, which requires the expected posterior mean to equal the prior mean. In the present binary-state setting, this condition is both necessary and sufficient for a distribution of posterior means to be induced by a feedback experiment.

Conversely, any distribution satisfying these constraints can be implemented by a signal structure. To see this, let $q_j = ( x_j - L ) / ( H - L )$ and define
\begin{equation}
    \sigma(j| H)=\frac{ F_jq_j}{p},
    \qquad
    \sigma(j| L)=\frac{ F_j(1-q_j)}{1-p}.  \label{eq_implement}
\end{equation}
Bayes plausibility ensures that both sets of conditional probabilities sum to one.

Finally, we define the net value of a feedback design relative to uninformative feedback.
The net value of feedback $F$ relative to uninformative feedback is given by
\begin{equation}
    \Delta_W ( F , \mu  ) \coloneqq \E_F [ W ( X ) ] - W ( \mu ) .
\end{equation}
Thus, an informative feedback experiment inducing $F$ is \emph{beneficial} at prior $\mu$ if $\Delta_W(F,\mu) > 0$. 
We use this measure to characterize when informative feedback is beneficial and to analyze the optimal feedback design.


\section{Benchmark Case: Constant Number of Messages}


To illustrate the role of changes in the equilibrium number of messages, consider first the benchmark case in which the equilibrium number of messages is constant for all $x \in [ L , H ]$.
In this case, the optimal feedback policy admits a simple characterization.

Suppose that $N(x) = N \ge 1$ for every $x\in[L,H]$.
Then, any feedback experiment, whether informative or uninformative, induces a continuation equilibrium with the same number of messages $N$. 
Hence, by using \eqref{eq_wn}, the value of feedback $F$ is
\begin{align*}
      \Delta_W(F,\mu)=&-\frac{1}{12N^2}+k_N\E[X^2]-\left(-\frac{1}{12N^2}+k_N\mu^2\right)  \\
      =&  k_N\var_F(X) .
\end{align*}

Thus, the coefficient $k_N$ measures the change in the designer's expected payoff per unit increase in the variance of posterior means, holding the number of messages fixed at $N$.
Its first component, $\alpha_R(2-\alpha_R)>0$, captures the benefit of reducing uncertainty about $\theta$, which allows decisions to better reflect the employee's potential; we refer to this as the \emph{attribute effect}. 
Its second component, $-d^2(N^2-1)/3\leq 0$, captures the increase in expected communication loss about $\omega$; we refer to this as the \emph{communication effect}.
Thus, $k_N$ represents the net benefit of informative feedback, balancing improved decisions based on $\theta$ against less effective communication about $\omega$.

Thus, when $k_N > 0$, the attribute effect outweighs the communication effect, and informative feedback raises the designer's expected payoff. Hence, full disclosure is optimal. Conversely, when $k_N < 0$, the communication effect dominates the attribute effect, so informative feedback reduces the designer's expected payoff.
Hence, no disclosure is optimal.
When $k_n=0$, every feedback experiment yields the same value. 

This leads to two important observations. First, either full disclosure or no disclosure is strictly optimal; in other words, partial disclosure is never strictly optimal. Second, because $k_N$ is independent of $p$, the optimal feedback design is independent of the probability of $\theta$ being high (i.e., the distribution of employees' potential). 

However, this simple characterization does not generally extend to the case where $N(x)$ varies with $x$.
In the general case, $W$ is piecewise quadratic and exhibits upward kinks at the thresholds where the equilibrium number of messages changes.
Consequently, $\Delta_W$ can no longer be summarized solely by the variance of posterior means. 
Characterizing the optimal feedback policy therefore requires a more careful analysis of the concave envelope of $W$, to which we now turn.

\section{Main Analysis}
\label{sec: main analysis}
We turn to our main analysis. While the previous analysis assumed that $N(x)$ is constant (i.e., there is no $n\geq 2$ such that $t_n\in[L, H]$), the following analysis includes other cases. This section is organized as follows. Section \ref{sec: MA-preliminaries} provides the geometric expression of the designer's problem. Section 5.2 examines when informative feedback is beneficial, and Section 5.3 derives a necessary and sufficient condition for full disclosure to be optimal. Section 5.4 characterizes optimal feedback policies when full disclosure is suboptimal. Finally, Section 5.5 summarizes the results and discusses their managerial implications.

\subsection{Preliminaries}
\label{sec: MA-preliminaries}


The designer's problem admits a simple geometric interpretation.
Suppose an experiment induces two posterior means $a<\mu<b$.
Bayes plausibility uniquely determines their probabilities as
$(b-\mu)/(b-a)$ and $(\mu-a)/(b-a)$, respectively.
The resulting expected payoff equals the height, at $\mu$,
of the straight line connecting $(a,W(a))$ and $(b,W(b))$.
By contrast, no disclosure yields $W(\mu)$.

This observation leads to the following characterization.
Let $\cav W$ denote the \emph{concave envelope} of $W$; that is, the smallest concave function lying weakly above $W$.

\begin{lemma}[Concavification]
    \label{lem:concavification}
    The designer's optimal transformed payoff is
    \begin{equation}\label{eq:concavification}
    \cav W(\mu)
    =
    \max_{L\le a\le\mu\le b\le H}
    \left\{
    \frac{b-\mu}{b-a}W(a)
    +
    \frac{\mu-a}{b-a}W(b)
    \right\},
    \end{equation}
    where $a=b=\mu$ yields $W(\mu)$.
    In particular, an optimal feedback experiment exists that induces at most two posterior means.
\end{lemma}

Thus, the standard technique in Bayesian persuasion is applied to our situation \citep{kamenica2011bayesian}.
Lemma \ref{lem:concavification} tells us that on each maximal interval where $V > W$, $V$ is the straight line connecting the contact points at the two endpoints.
If $V = W$, an uninformative feedback is optimal.

The remaining task is therefore to characterize the concave envelope of $W$.
As we show below, the upward kinks established in Lemma~\ref{lem:envelope} play a central role in determining the structure of the optimal feedback policy.

\subsection{When is Informative Feedback Beneficial?}


In what follows, we examine the general case in which the number of equilibrium messages varies with the posterior mean. The value of feedback can be written as
\begin{equation}
    \Delta_W(F,\mu)
    =
    \underbrace{\alpha_R(2-\alpha_R)\var_F(X)}_{\rm attribute\ effect}
    -
    \underbrace{
    \E_F\left[
        \mathcal L_{N(X)}(X)
        -
        \mathcal L_{N(\mu)}(\mu)
    \right]}_{\rm communication\ effect},
\end{equation}
where $\mathcal L_n(x)$ denotes the communication loss at posterior mean $x$ when the number of messages is $n$, as defined in \eqref{eq: communication loss}.

As in the benchmark case, the attribute effect captures the benefit of reducing uncertainty about $\theta$ and is strictly positive for informative feedback. The communication effect, in contrast, captures how feedback changes the expected loss from imperfect communication about $\omega$. By Bayes plausibility, informative feedback induces posterior means both above and below the prior mean $\mu$. A lower posterior mean weakens the sender--receiver preference conflict and improves communication, whereas a higher posterior mean has the opposite effect. While the communication effect is nonpositive when $N$ is constant, its sign is generally ambiguous when $N$ varies, depending on whether the improvement in communication at lower posterior means outweighs the deterioration at higher posterior means.

To understand the communication effect, we decompose it into two components. The \emph{fixed-message effect} captures the change in expected communication loss arising from variation in posterior means, holding the number of messages fixed at $N(\mu)$. The \emph{message-adjustment effect} captures the additional change arising from adjustments in the equilibrium number of messages. Using this decomposition, we obtain
\begin{equation}
\begin{aligned}
    \Delta_W(F,\mu)
    =&\underbrace{
        \alpha_R(2-\alpha_R)\var_F(X)
    }_{\rm attribute\ effect}-\underbrace{
        \frac{d^2(N(\mu)^2-1)}{3}\var_F(X)
    }_{\rm fixed\text{-}message\ effect}+\underbrace{
        \E_F[W(X)-W_{N(\mu)}(X)]
    }_{\rm message\text{-}adjustment\ effect}\\
    =&\underbrace{\ k_{N(\mu)}\var_F(X)}_{\rm attribute \ effect \ + \ fixed\text{-}message\ effect}
    +\underbrace{\E_F[W(X)-W_{N(\mu)}(X)]}_{\rm message\text{-}adjustment\ effect}.
\end{aligned}
\end{equation}
The first two terms together represent the net value of feedback when the number of messages is held fixed, as in the benchmark case.
The last term captures the additional value arising from endogenous adjustments in the number of messages.

To illustrate this decomposition, consider $\alpha_R=1$, $d=4/3$, $L=0.05$, and $H=0.20$. Suppose that the prior mean is $\mu=0.125$ and the designer fully discloses the attribute. The equilibrium number of messages is two without feedback, but increases to three following the lower posterior mean and decreases to one following the higher posterior mean.

Table \ref{tab:feedback_decomposition} reports the contribution of each effect. The attribute effect is positive, but is more than offset by the negative fixed-message effect. Nevertheless, the message-adjustment effect is sufficiently large to make the net value of feedback positive. In particular, message adjustment accounts for 53.2\% of the total positive gains from feedback, while the fixed-message effect offsets 83.2\% of these gains. Thus, endogenous adjustments in communication can make informative feedback beneficial even when feedback would be detrimental with a fixed number of messages.

\begin{table}[t]
\centering
\caption{Decomposition of the value of feedback}
\label{tab:feedback_decomposition}
\begin{tabular}{lrr}
\hline\hline
Effect & Value & Share of net gain (\%) \\
\hline
Attribute effect
    & $0.005625$ & $279.3$ \\
Fixed-message effect
    & $-0.010000$ & $-496.6$ \\
Message-adjustment effect
    & $0.006389$ & $317.2$ \\
\hline
Net value of feedback
    & $0.002014$ & $100.0$ \\
\hline\hline
\end{tabular}

\medskip
\begin{minipage}{0.95\linewidth}
\footnotesize
\textit{Notes:}
$\alpha_R=1$, $d=4/3$,
$L=0.05$, and $H=0.20$.
The prior mean is $\mu=0.125$.
Full disclosure induces $X=0.05$ and $X=0.20$
with equal probabilities.
The equilibrium numbers of messages are
$N(\mu)=2$, $N(0.05)=3$, and $N(0.20)=1$.
Shares are calculated relative to the net
value of feedback.
\end{minipage}
\end{table}

In summary, for any posterior distribution $F$, the value of feedback is given by
\begin{equation}
    \Delta_W ( F , \mu ) =
    k_{ N ( \mu ) } \var_F ( X )
    + \E_F [ W ( X ) - W_{ N ( \mu ) } ( X ) ]
\end{equation}
Unlike in the benchmark case, therefore, the sign of $k_{N(\mu)}$ does not determine whether informative feedback is beneficial or characterize the optimal feedback experiment.
The designer must also take the message-adjustment effect into account.
Thus, in the general case, the optimal feedback experiment is determined by the interaction between the attribute and fixed-message effects, summarized by $k_{N(\mu)}$, and the message-adjustment effect.

Although endogenous changes in the number of messages complicate the evaluation of feedback, they do not necessarily make informative feedback less desirable. On the contrary, when the number of equilibrium messages varies across interior posterior means, informative feedback is strictly beneficial for some priors, as the following proposition shows.

\begin{proposition}[Beneficial feedback]
    \label{prop: sufficient condition for beneficial feedback}
    Suppose that there exists $n\geq 2$ such that $t_n\in(L,H)$.
    Then, there exists a nonempty open interval of priors for which an informative feedback experiment is beneficial.
\end{proposition}

\begin{figure}[t]
    \centering
    \includegraphics[width=.8\linewidth]{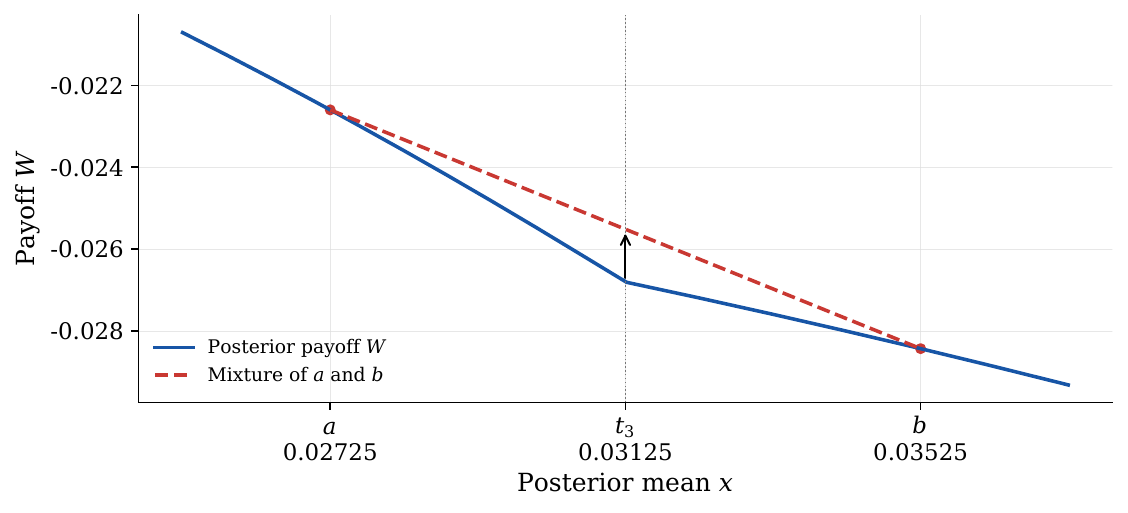}
    \caption{A small split around a threshold where the equilibrium message count falls improves the payoff. At the prior mean $t_3=0.03125$, no disclosure gives the blue curve. Equal probabilities on $a=t_3-0.004$ and $b=t_3+0.004$ preserve that mean and give the height of the red line. The arrow marks the gain. }\label{fig:kink}
\end{figure}

Figure~\ref{fig:kink} illustrates the key intuition.
At a threshold $t_n$, the function $W$ is continuous but has an upward kink by Lemma~\ref{lem:envelope}.
Therefore, a small mean-preserving spread around $t_n$ yields $\E_F[W(X)]>W(t_n)$, and hence generates a strictly positive net value of feedback at prior $\mu=t_n$.
In the figure, replacing the degenerate posterior at $t_3$ with a mixture of two nearby posterior means $a < t_3 < b$ raises the expected value of $W$ from its value at the kink to the value on the red chord.
Thus, uninformative feedback is strictly suboptimal at the threshold. This contrasts with the case where $N$ is constant, in which no disclosure can be optimal for all priors.

\subsection{Optimality of Full Disclosure}
\label{sec: full disclosure}
We now examine when full disclosure is optimal. Any partial disclosure experiment induces posterior beliefs about $\theta$. To assess whether full disclosure outperforms such experiments, consider an arbitrary posterior mean $x\in[L,H]$ and compare the payoff $W(x)$ with the expected payoff from fully revealing $\theta$ starting from that belief, denoted by $W^{FD}(x)$. This comparison is important because, if the latter is at least as high as the former for every $x$, inducing an interior posterior belief and stopping disclosure there can never be better than fully revealing the attribute from that belief.

Under full disclosure, the posterior mean becomes either $L$ or $H$. Thus,
\[
W^{FD}(x)=\frac{H-x}{H-L}W(L)+\frac{x-L}{H-L}W(H).
\]
Geometrically, $W^{FD}$ is the line segment connecting $(L,W(L))$ and $(H,W(H))$.

Therefore, the difference between $W^FD(x)$ and $W(x)$ can be decomposed as
\begin{align}
    W^{FD} ( x ) - W ( x )
    &= \alpha_R ( 2 - \alpha_R ) ( x - L ) ( H - x )
    \\
    &\quad
    - \left[
      \frac{ H - x }{ H - L } \mathcal L_{ N ( L ) } ( L ) + \frac{ x - L }{ H - L } \mathcal L_{ N ( H ) } ( H ) - \mathcal L_{ N ( x ) } ( x )
    \right]
    .
    \label{eq:fulltradeoff}
\end{align}
The first term is the attribute effect from fully revealing $\theta$.
It is strictly positive for every interior belief since $0<\alpha_R<2$.
The second term is the communication effect, which measures how full disclosure changes expected communication loss relative to leaving the belief unchanged.
A positive communication effect offsets the attribute effect, whereas a negative one reinforces it.

It can be shown that full disclosure is optimal if and only if
\begin{equation}
    W^{FD}(x)\geq W(x)
    \qquad\text{for all }x\in[L,H]. \label{eq_full}
\end{equation}
The following proposition translates this condition into a necessary and sufficient condition for full disclosure to be optimal.

\begin{proposition}[Optimality of full disclosure]
    \label{prop:full}
    Full disclosure is optimal if and only if
    \begin{equation}
        \label{eq:fullthreshold}
        \alpha_R(2-\alpha_R)\ge
        \sup_{L<x<H}
        \frac{\displaystyle
        \frac{H-x}{H-L} \mathcal L_{ N ( L ) } (L)
        +\frac{x-L}{H-L} \mathcal L_{ N ( H ) } (H) - \mathcal L_{ N ( x ) } ( x )
        }{(x-L)(H-x)}.
    \end{equation}
    Otherwise, it is strictly suboptimal
    for every interior prior.
\end{proposition}

Proposition~\ref{prop:full} features the trade-off between the attribute effect and the communication effect.
The left-hand side of \eqref{eq:fullthreshold} captures the attribute effect of full disclosure, whereas the right-hand side captures its adverse effect on communication.
Thus, full disclosure is optimal when the attribute effect is sufficiently strong relative to the communication effect.
In particular, holding $d,L,H$ fixed, the threshold does not depend on $\alpha_R$, whereas the attribute-related benefit is governed by $\alpha_R(2-\alpha_R)$. An important caveat is that because \eqref{eq_full} must hold for any $x$, the prior $p$ is irrelevant for the condition. Indeed, condition \eqref{eq:fullthreshold} is independent of $p$; that is, if full disclosure is optimal, it is always the case irrespective of the prior belief about $\theta$.

\begin{figure}[t]
\centering
\includegraphics[width=.94\linewidth]{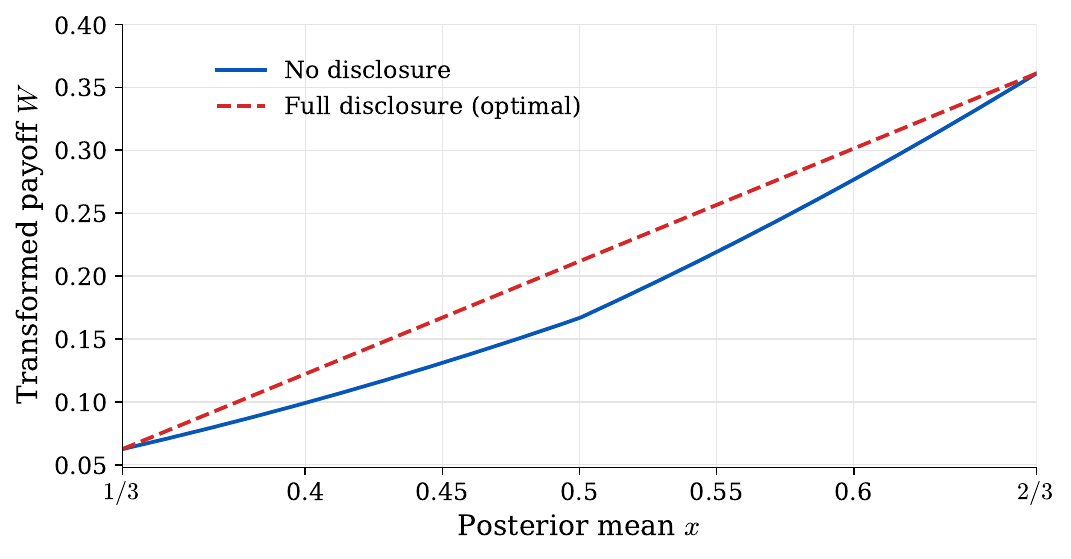}
\caption{Full disclosure can be optimal even when the equilibrium number of messages varies. Here $\alpha_R=1$, $\alpha_S=3/2$, $L=1/3$, and $H=2/3$. The number of messages decreases from two to one at $x=1/2$.}
\label{fig:full}
\end{figure}

To illustrate that full disclosure can be optimal, let $\alpha_R=1$, $\alpha_S=3/2$, $L=1/3$, and $H=2/3$.
In this case, the equilibrium number of messages is
\[
N(x)=
\begin{cases}
2 & \text{if } x<1/2,\\
1 & \text{if } x\geq 1/2.
\end{cases}
\]
The designer's transformed payoff is
\[
W(x)=\max\left\{
-\frac{1}{48}+\frac{3}{4}x^2,\,
-\frac{1}{12}+x^2
\right\}.
\]
Since both quadratic functions are strictly convex, $W$ is strictly convex on $[L,H]$.
Consequently, $W^{FD}(x)>W(x)$ for every $x\in(L,H)$, implying that full disclosure is optimal for every interior prior.
Figure~\ref{fig:full} illustrates this result.

\paragraph{Small and large sender--receiver preference differences:} To see the effect of sender-receiver preference differences on condition (\ref{eq:fullthreshold}), we fix $\alpha_R,L,H$ and vary $\alpha_S$, so that $d$ changes. This comparison changes the number of messages that can be sustained. It need not have a monotone effect on the desirability of full disclosure. Nevertheless, as a corollary of the proposition, the two extremes can be characterized analytically.

\begin{corollary}[Full disclosure with small or large communication conflict]
\label{prop:extremes}
The following holds:
\begin{enumerate}
    \item Full disclosure is optimal for sufficiently small $d>0$.
    \item Full disclosure is optimal whenever $dL\ge1/4$.
\end{enumerate}
\end{corollary}

For small conflict, communication is sufficiently accurate that the value of attribute information makes full disclosure optimal. For large conflict, communication already conveys no state information at any posterior, so attribute revelation cannot damage it further. Therefore, in these two extreme cases, full disclosure is optimal.\footnote{These conditions are sufficient but not necessary.
In general, the values of $d$ for which full disclosure is
strictly suboptimal need not form a single interval.
For example, fix $\alpha_R=1/10$, $L=1/20$, and $H=1/5$.
Full disclosure is optimal at $d=0.3$, $0.5$, and $2$, but is
strictly suboptimal at $d=0.4$ and $1$.}
This implies that the effect of conflict of interests on full disclosure is non-monotonic.

\subsection{Optimal Feedback Design}
\label{sec:nonfull}

We now turn to the case in which condition \eqref{eq:fullthreshold} fails, so that fully informative feedback is strictly suboptimal for every interior prior. The designer's problem is therefore $\max_F \Delta_W ( F , \mu )$.
We first investigate when partial disclosure is optimal and then characterize the structure of the optimal feedback experiment across priors.

We begin by establishing the optimality of partial disclosure.
For a given prior $\mu$, partial disclosure is strictly optimal if there exists a feedback experiment $F$ such that
\begin{equation}
    \Delta_W(F,\mu) > \max\{0,\Delta_W(F^{FD},\mu)\}.
\end{equation}
Thus, partial disclosure must outperform both uninformative feedback, whose net value is zero, and fully informative feedback. By combining Proposition \ref{prop: sufficient condition for beneficial feedback} and \ref{prop:full}, we can easily show the following as a corollary:
\begin{corollary}[Optimality of partial disclosure]
    \label{prop: optimality of partial disclosure}
    Suppose that condition \eqref{eq:fullthreshold} fails.
    Partial disclosure is strictly optimal for a nonempty open interval of interior priors if and only if there exists $n\geq 2$ such that $t_n\in(L,H)$.
\end{corollary}

This corollary shows that, when fully informative feedback is suboptimal, the existence of an interior communication threshold $t_n$ is both necessary and sufficient for partial disclosure to be optimal for some priors.
An interior threshold allows the equilibrium number of messages to adjust to the posterior mean, creating a message-adjustment effect.
Without such an adjustment, the problem reduces to the benchmark case, in which the suboptimality of fully informative feedback implies the optimality of uninformative feedback.

We next characterize the structure of the optimal feedback experiment across priors.
Recall that the net value of feedback can be decomposed into the message-adjustment effect and the net effect of inducing dispersion in posterior means while holding the number of messages fixed at $N(\mu)$, with the latter governed by $k_{N(\mu)}$.
Thus, the optimal feedback experiment generally balances these two forces.
Depending on the sign of $k_{N(\mu)}$, dispersion in posterior means may either reinforce or offset the gain from the message-adjustment effect.
Hence, the designer must choose posterior means that generate a sufficiently large message-adjustment effect while taking into account the effect of dispersion for a given number of messages. 
Thus, the optimal feedback experiment depends not only on the amount of dispersion it induces, but also on the location of the induced posterior means relative to the communication thresholds.

Despite the discrete changes in the equilibrium number of messages across communication thresholds, the optimal feedback policy admits a simple structure across priors.
The next proposition bounds the number of regions in which uninformative feedback is optimal and characterizes the posterior supports used in the remaining regions.

\begin{proposition}[Structure of optimal feedback]
    \label{thm:two}
    Suppose condition (\ref{eq:fullthreshold}) fails.
    The set of interior priors at which no disclosure is either one or two intervals. Consequently, an optimal feedback
    policy has at most five successive regions as $p$ increases:
    \begin{enumerate}
    \item \emph{One no-disclosure interval (three-region structure).}
    There exist $0\le p_-\le p_+\le1$ such that no disclosure is
    optimal for $p\in[p_-,p_+]\cap(0,1)$ and strict partial disclosure
    is optimal elsewhere. An optimal experiment induces the following
    posterior-probability supports:
    \begin{equation}\label{eq:optimal-three-supports}
    \{0,p_-\}\quad(0<p<p_-),\qquad
    \{p\}\quad(p_-\le p\le p_+),\qquad
    \{p_+,1\}\quad(p_+<p<1).\notag
    \end{equation}
    
    \item \emph{Two no-disclosure intervals (five-region structure).}
    There exist $0\le p_1\le p_2<p_3\le p_4\le1$ such that no
    disclosure is optimal for
    $p\in([p_1,p_2]\cup[p_3,p_4])\cap(0,1)$,
    and strict partial disclosure is optimal elsewhere. An optimal
    experiment induces the posterior-probability supports
    \begin{equation}\label{eq:optimal-five-supports}
    \begin{cases}
    \{0,p_1\}, & 0<p<p_1,\\
    \{p\}, & p_1\le p\le p_2,\\
    \{p_2,p_3\}, & p_2<p<p_3,\\
    \{p\}, & p_3\le p\le p_4,\\
    \{p_4,1\}, & p_4<p<1.
    \end{cases} \notag
    \end{equation}
    \end{enumerate}
\end{proposition}

This proposition provides a complete characterization
of how the optimal feedback policy varies with the prior.
Its first implication is that whenever full disclosure is
suboptimal, no disclosure must be optimal for a nonempty
interval of priors. Thus, withholding all information is
not merely an exceptional possibility that arises at
isolated parameter values. At the same time, the proposition
shows that there can be at most two separate no-disclosure
intervals, so the optimal policy has at most five regions
as the prior varies.

Combined with Proposition~\ref{prop: optimality of partial disclosure}, this result
also identifies when both no disclosure and partial
disclosure arise. If the equilibrium number of messages
changes at some $t_n\in(L,H)$, both policies are optimal
on nonempty open intervals of priors. Thus, even with
fixed preferences and attribute values, the optimal
feedback policy can differ depending on the prior
probability of the high attribute. Conversely, if
$N(x)$ remains constant, no disclosure is optimal
for every prior whenever full disclosure is suboptimal.

The proposition further characterizes the information
provided by an optimal feedback policy. In the
three-region case, the optimal policy follows a simple
pattern as $p$ increases: partial disclosure, no
disclosure, and partial disclosure. In the first region,
an optimal experiment induces posterior probabilities
$\{0,p_-\}$. One signal perfectly identifies the low
attribute, while the other leaves uncertainty about
the attribute. In the last region, the optimal support
is instead $\{p_+,1\}$: one signal perfectly identifies
the high attribute, while the other remains inconclusive.
Thus, the type that can be conclusively identified
switches as the prior increases.

The five-region case exhibits a richer pattern.
In addition to the two outer partial-disclosure regions,
there is a middle partial-disclosure region separating
two no-disclosure intervals. Importantly, the optimal
posterior support in this middle region is
$\{p_2,p_3\}$, where both posterior probabilities
can be strictly between zero and one. Unlike the
outer regions, neither signal needs to reveal the
attribute perfectly. This shows that optimal partial
disclosure need not take the form of conclusively
identifying one attribute type.

\begin{figure}[t]
    \centering
    \begin{minipage}[t]{.48\linewidth}
        \vspace{0pt}
        \centering
        \includegraphics[width=\linewidth]{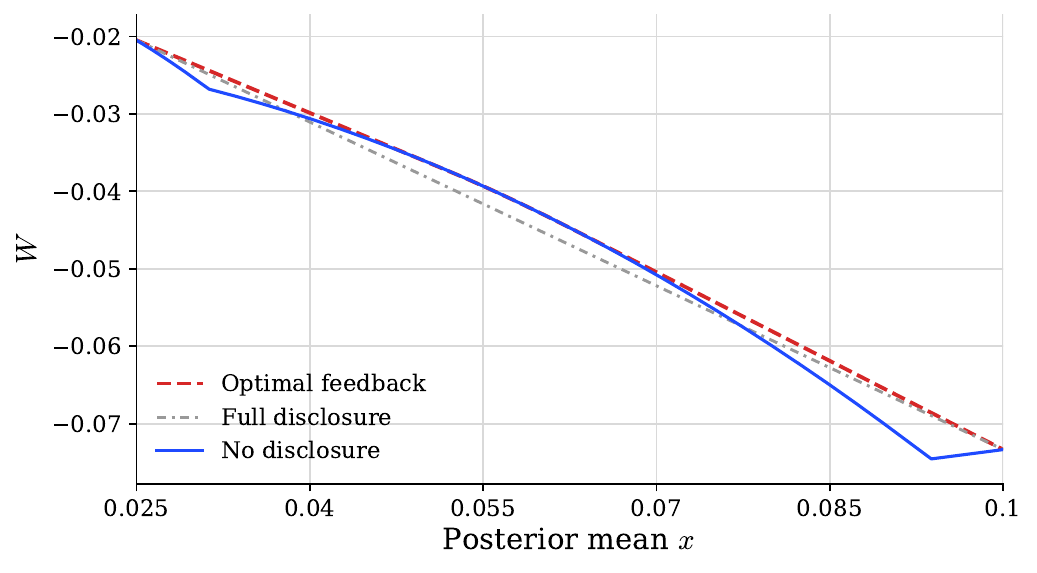}\par
        \smallskip
        \textbf{(i)} Three-region structure
    \end{minipage}\hfill
    \begin{minipage}[t]{.48\linewidth}
        \vspace{0pt}
        \centering
        \includegraphics[width=\linewidth]{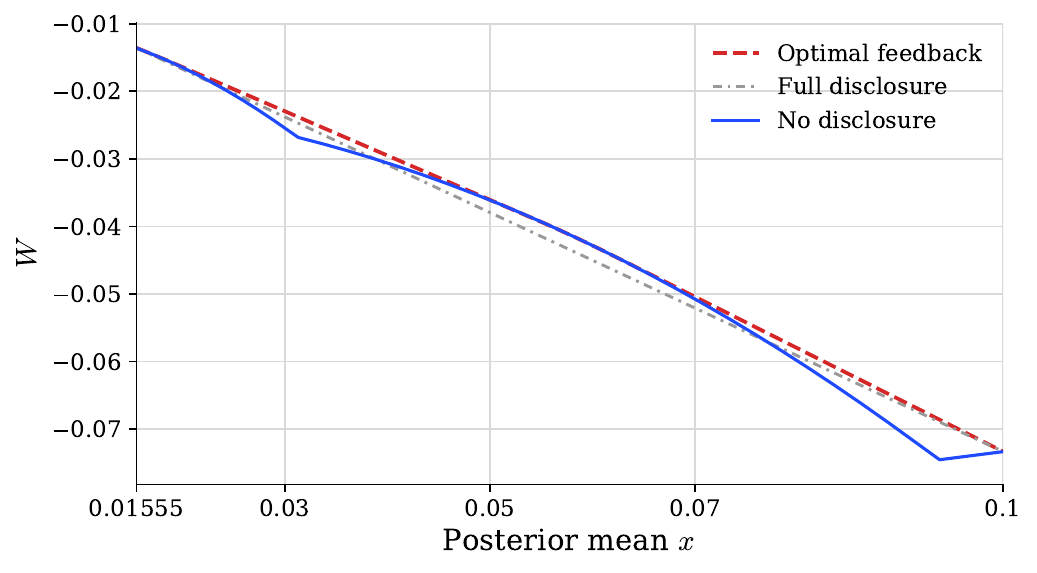}\par
        \smallskip
        \textbf{(ii)} Five-region structure
    \end{minipage}

    \caption{Three- and five-region structures of optimal disclosure.
    In both panels, the solid blue curve is the no-disclosure payoff,
    and the dashed red curve is its concave envelope.
    In panel (a), the curves coincide on $[p_-,p_+]$.
    The supporting chords to the left and right induce posterior supports
    $\{0,p_-\}$ and $\{p_+,1\}$, respectively.
    In panel (b), the curves coincide on two intervals.
    From left to right, the three supporting chords induce posterior supports
    $\{0,p_1\}$, $\{p_2,p_3\}$, and $\{p_4,1\}$.}
    \label{fig: optimal disclosure structure}
\end{figure}

Another notable property is that the optimal posterior
support remains fixed within each partial-disclosure
region. As the prior changes, the designer adjusts
the probabilities of inducing these posteriors rather
than the posterior beliefs themselves. Specifically,
for an optimal support $\{a,b\}$ expressed in posterior
probabilities, with $0\le a<p<b\le1$,
Bayes plausibility gives
\begin{equation}\label{eq:optimal-signal-weights}
\Pr(s_a)=\frac{b-p}{b-a},\qquad
\Pr(s_b)=\frac{p-a}{b-a}.
\end{equation}
The corresponding conditional signal probabilities
follow from \eqref{eq_implement}.
Hence, the same pair of posterior beliefs can be
induced optimally for an entire interval of priors,
with only the frequencies of the two signals changing.

Panel (i) of Figure~\ref{fig: optimal disclosure structure} illustrates the three-region
structure. Let $\alpha_R=1$, $\alpha_S=11/3$,
$L=1/40$, and $H=1/10$. No disclosure is optimal
for $p\in[p_-,p_+]$, where
$p_-=\sqrt{4/33}\simeq0.3482$ and
$p_+=1-\sqrt{124/495}\simeq0.4995$.
In this interval, the optimal-payoff curve coincides
with the no-disclosure payoff. Outside it, the optimal
payoff is given by a straight line connecting an
endpoint to an interior contact point. For low priors,
the designer sometimes reveals that the employee has
low potential, whereas for high priors, the designer
sometimes reveals that the employee has high potential.
Full disclosure is strictly suboptimal throughout.

Panel (ii) of Figure~\ref{fig: optimal disclosure structure} illustrates that the
five-region structure can also arise. Consider
$\alpha_R=1$, $d=8/3$, $L=311/20000$, and $H=1/10$.
In this example, there are two distinct no-disclosure
intervals, separated by a middle partial-disclosure
region. As the prior increases, the optimal policy
alternates between partial disclosure and no disclosure
five times in the sense of five successive policy regions.

The figure also highlights an important difference
between the outer and middle partial-disclosure
regions. In the outer regions, an optimal experiment
has one fully revealing signal. In the middle region,
however, both posterior beliefs are interior.
The designer therefore benefits from providing
information about employee potential without ever
revealing it completely. The concave envelope
coincides with the no-disclosure payoff on two
separate intervals and lies strictly above it
elsewhere. Thus, the optimal degree and form of
feedback need not vary monotonically with the prior.

\paragraph{Proof sketch of Proposition \ref{thm:two}}

The proof relies on the geometry of the concave envelope of $W$.
By Lemma~\ref{lem:envelope}, $W$ has an upward kink at every communication threshold $t_n$.
Such a threshold cannot be an interior contact point between $W$ and its concave envelope. Hence, any interval on which uninformative feedback is optimal must lie within a region where the equilibrium number of messages is constant.
On the complementary regions, partial disclosure is optimal and represented by a binary experiment.
The locations of its interior posterior means solve the maximization problem for $\Delta_W$: they balance the gain from the message-adjustment effect against the effect of inducing dispersion in posterior means while holding the number of messages fixed.
Geometrically, this balance implies that the line connecting two interior contact points must support $W$ at both points, yielding the tangency condition
\begin{equation}
    W'(a)=W'(b)=\frac{W(b)-W(a)}{b-a}.
\end{equation}

The key restriction comes from the ordering of these supporting slopes across message regions.
The intuition is closely related to the Crawford--Sobel model.
An equilibrium with more messages permits more informative communication when the sender--receiver conflict is small, but its value is also more sensitive to an increase in the conflict.
Thus, as the posterior mean rises and the conflict becomes larger, the reduction in value is greater for an equilibrium with more messages than for one with fewer messages.
In the proof, this ordering is reflected in $\frac{-k_n}{n^2}>\frac{-k_m}{m^2}$ for $n > m$, which generates the corresponding monotonicity in the slopes of the supporting lines between contact regions.

If there were three distinct regions in which uninformative feedback were optimal, the supporting lines connecting successive contact regions would therefore have slopes ordered in the opposite direction from that required by concavity.
Hence, the contact set can contain at most two nondegenerate intervals. This restriction is what gives the optimal feedback policy its simple structure despite the discrete changes in the equilibrium number of messages across communication thresholds.



\subsection{Summary and Managerial Implications}\label{sec:managerial}

\begin{figure}[t]
\centering
\includegraphics[width=.96\linewidth]{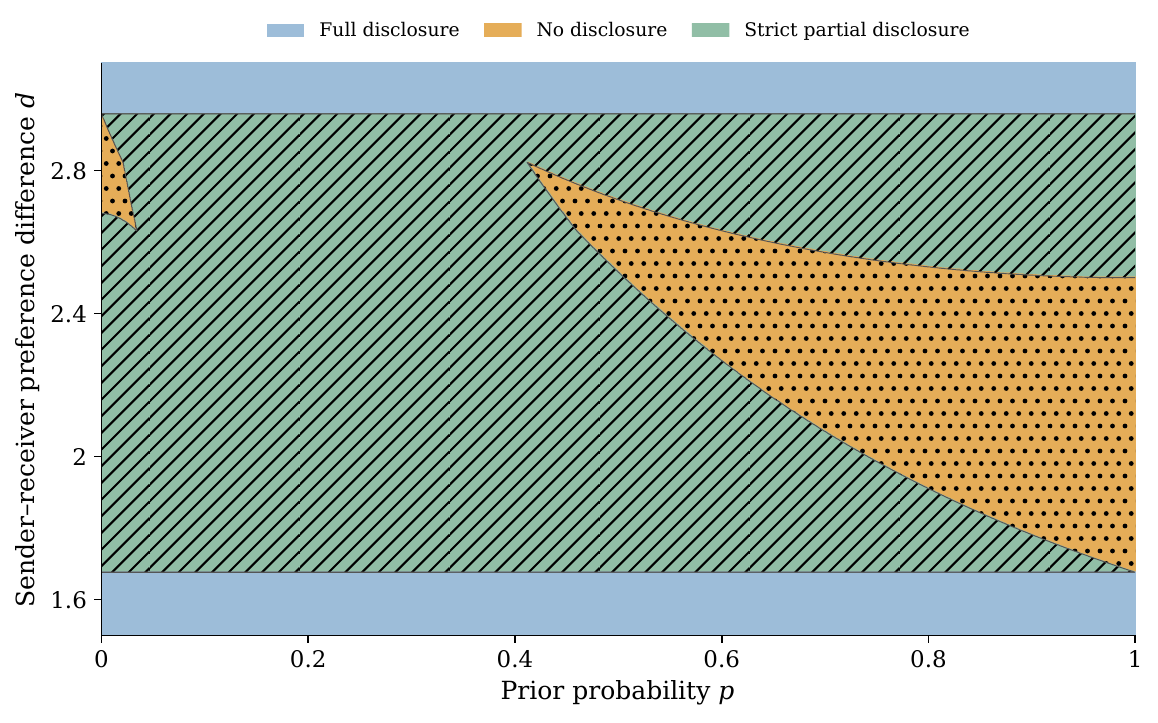}
\caption{Optimal feedback policies as the prior probability of high
potential $p$ and the sender--receiver preference difference $d$ vary.
Parameters are $\alpha_R=1$, $L=311/20000$, and $H=1/10$.}
\label{fig:policy-map}
\end{figure}

Our analysis highlights a trade-off in the design of employee feedback.
Information about an employee's potential helps the manager tailor the
assignment to the employee, but it also changes the employee's incentives
to communicate private information about project-specific circumstances.
Figure~\ref{fig:policy-map} summarizes how this trade-off shapes the optimal
feedback policy as the prior probability $p$ of high potential and the
sender--receiver preference difference $d$ vary. The figure fixes
$\alpha_R=1$, $L=311/20000$, and $H=1/10$; hence the manager's preferences
are aligned with the firm's. 

The figure brings together two aspects of our characterization.
First, whether full disclosure is optimal does not depend on $p$
(Proposition~\ref{prop:full}). Accordingly, the full-disclosure regions
extend across the entire horizontal axis. In the illustrated parameter
range, full disclosure is optimal at relatively low and high values of $d$,
but not for an intermediate range. Second, when full disclosure is
suboptimal, there is always a nondegenerate interval of priors for which
no disclosure is optimal (Proposition~\ref{thm:two}). If the equilibrium
number of communication messages changes with the posterior belief,
strict partial disclosure is also optimal on an open interval
(Corollary~\ref{prop: optimality of partial disclosure}). For any fixed $d$, the resulting policy
can switch between no disclosure and partial disclosure, but it has at
most five regions. The figure includes the five-region configuration
at $d=8/3$, in which two separate no-disclosure intervals are separated
by a middle partial-disclosure interval.

These results provide the following three managerial implications for feedback design.

\paragraph{More informative evaluations need not improve decisions:}
Even with a manager whose objectives coincide with those of the firm,
full disclosure can be suboptimal. The reason is that evaluations do more
than inform the manager about the employee's potential: because feedback
is public, they also affect the employee's preferred assignment.
A favorable signal raises the posterior mean $x$ and therefore widens
the sender--receiver preference difference in the continuation game,
$dx$. This can reduce the informativeness of cheap-talk messages about
$\omega$. The firm may consequently prefer to sacrifice some precision
about employee potential in order to preserve communication about the
circumstances of a particular project. The relevant design objective is
thus not the informativeness of feedback in isolation, but its effect
on the quality of the final assignment decision.

\paragraph{Preference conflict changes the value of transparency:}
Figure~\ref{fig:policy-map} illustrates that the desirability of full
feedback need not be monotone in $d$. More generally, the two extremes
help explain why. When preferences are closely aligned, subsequent
communication is relatively informative, so revealing potential has
little cost in terms of lost information about the project. If the
conflict is sufficiently large that only one cheap-talk message is
feasible at every posterior, communication is already uninformative,
and revealing potential cannot worsen it further. At intermediate
conflict levels, disclosure may instead change the informativeness
of communication, making partial or no disclosure attractive. Thus,
the optimal evaluation system should reflect the incentives governing communication after
evaluations are announced.

\paragraph{The appropriate evaluation standard depends on the workforce:}
The optimal partial-disclosure policy also depends on the prevalence of
high-potential employees. In a partial-disclosure region with a
low prior $p$, an optimal experiment can induce posterior probabilities
$\{0,p_-\}$. One assessment conclusively identifies low potential,
while the other raises the assessment of potential without establishing
high potential with certainty. Interpreted as a rating system, this
resembles a relatively lenient standard for favorable evaluations:
unfavorable evidence is decisive, whereas favorable evaluations are less
discriminating. At higher priors, an optimal experiment can instead
induce $\{p_+,1\}$. Here a favorable evaluation conclusively identifies
high potential, while the other assessment remains inconclusive. This
resembles a more selective standard for a favorable evaluation. These
interpretations concern the evidentiary content of evaluations, not
rating quotas or comparisons among employees.

The five-region case adds a further possibility. For intermediate
priors, the optimal experiment can induce two strictly interior
posteriors, so neither evaluation conclusively identifies the employee's
type.  In addition, for the priors
in a no-disclosure region, the firm is better off withholding the
evaluation entirely, even though informative signals about potential
are technologically feasible.

\section{Concluding Remarks}
Employee assessments provide valuable information for managerial decisions, but revealing this information may also change employees' incentives to communicate what they know. This paper examined how firms should design feedback about employee potential when such feedback affects subsequent strategic communication. In particular, we investigated whether providing more informative evaluations improves managerial decisions and how the optimal feedback policy depends on the characteristics of employees and their incentives.

We developed a model that combines information design with cheap talk. A firm designs public feedback about an employee's potential, after which the employee communicates private information about project-specific circumstances to a manager. We showed that revealing employee potential has two opposing effects. While it helps the manager make better-informed decisions, it can also increase the conflict of interest between the employee and the manager, thereby reducing the informativeness of subsequent communication. Consequently, full disclosure can be suboptimal even when the manager's preferences are perfectly aligned with those of the firm.

We then characterized the optimal feedback policy. First, we derived a necessary and sufficient condition for full disclosure to be optimal and showed that full disclosure is optimal when the preference conflict is sufficiently small or sufficiently large. Second, when full disclosure is suboptimal, no disclosure is optimal for some prior beliefs, while partial disclosure can also be optimal. Importantly, the optimal policy need not vary monotonically with the prevalence of high-potential employees. We showed that there can be at most two separate no-disclosure regions, implying at most five successive policy regions. Moreover, the optimal evaluation standard can shift from relatively lenient to stringent as high-potential employees become more prevalent. These results suggest that firms should design evaluations not only to convey information about employees' potential but also to preserve the informativeness of subsequent communication.

\FloatBarrier

\appendix
\section{Omitted Proofs}

\subsection{Proof of Lemma~\ref{lem:capacity}}


Fix a posterior mean $x$. By the equilibrium characterization of \citet{crawford1982strategic}, it suffices to consider interval-partition equilibria.

Consider an equilibrium with $n\ge2$ positive-length intervals, with boundaries
\[
0=\omega_0<\omega_1<\cdots<\omega_n=1.
\]
Conditional on receiving the message corresponding to interval $i$, the receiver's optimal action is
\[
y_i=\frac{\omega_{i-1}+\omega_i}{2}+\alpha_Rx,
\]
as stated in \eqref{eq:receiver}.

At each interior boundary $\omega_i$, sender indifference between adjacent actions requires
\[
\omega_i+\alpha_Sx=\frac{y_i+y_{i+1}}2.
\]
Substituting the receiver's actions yields
\begin{equation}\label{eq:app-recurrence}
\omega_{i+1}-\omega_i
=\omega_i-\omega_{i-1}+4dx.
\end{equation}
Writing $\ell_i=\omega_i-\omega_{i-1}$, we obtain
\[
\ell_{i+1}=\ell_i+4dx.
\]
Since $\sum_{i=1}^n\ell_i=1$, this implies
\[
\ell_i=\frac1n+2dx(2i-n-1).
\]
Summing the interval lengths gives
\[
\omega_i=\frac{i}{n}+2dx\,i(i-n).
\]

Because $d>0$ and $x>0$, the interval lengths are strictly increasing in $i$. Hence, all intervals have positive length if and only if $\ell_1>0$, which is equivalent to
\[
2dx\,n(n-1)<1.
\]
When this condition holds, the receiver's actions are strictly increasing, and the sender's quadratic preferences ensure that adjacent indifference implies global incentive compatibility. Thus, the constructed partition constitutes an equilibrium.

Finally, a single-message equilibrium always exists because the receiver can choose the optimal action based on the prior over $\omega$, independently of the sender's message.\qed

\subsection{Derivation of the Designer's Conditional Payoff}
\label{app: designer-conditional-payoff}

Let $\bar \omega_i = ( \omega_{ i - 1 } + \omega_i ) /2$.
The designer's conditional expected payoff given $s$ is
\begin{equation}
    \E [ u_0 | s ] =
    - \sum_{ i = 1 }^N \int_{ \omega_{ i - 1 } }^{ \omega_i } \left( \bar \omega_i + \alpha_R x - \omega - \theta \right)^2 d \omega .
\end{equation}
Observing that
\begin{equation}
    ( \bar \omega_i + \alpha_R x - \omega - \theta )^2
    = ( \bar \omega_i - \omega )^2 + 2 ( \bar \omega_i - \omega ) ( \alpha_R x - \theta ) + ( \alpha_R x - \theta )^2 .
\end{equation}
Since the expectation of $\omega$ at each interval is $\bar \omega_i$, we have $\E [ ( \bar \omega_i - \omega ) ( \alpha_R x - \theta ) | s ] = 0$.
Thus, we obtain $\E [ u_0 | s ] = - \E \left[ \left. \left( ( \omega_i + \omega_{ i - 1 } ) / 2 - \omega \right)^2 \right| s \right] - \E[(\alpha_Rx-\theta)^2| s]$.

Fix a posterior mean $x$ and a feasible $n$-message equilibrium. Since $\omega$ is uniform on $[0,1]$, the expected within-interval squared error is
\[
\mathcal L_n(x)=\sum_{i=1}^n\int_{\omega_{i-1}}^{\omega_i}
\left(\omega-\frac{\omega_{i-1}+\omega_i}{2}\right)^2d\omega
=\frac1{12}\sum_{i=1}^n\ell_i^3.
\]
Substituting \eqref{eq:lengths} and using
\[
\sum_{i=1}^n(2i-n-1)
=\sum_{i=1}^n(2i-n-1)^3=0,
\qquad
\sum_{i=1}^n(2i-n-1)^2=\frac{n(n^2-1)}3
\]
yields \eqref{eq: communication loss}. The cross term between the within-interval error and $\alpha_Rx-\theta$ has zero conditional expectation because $\omega$ and $\theta$ are independent conditional on $s$, and the within-interval error has mean zero. Finally, since $\theta\in\{L,H\}$,
\[
\theta^2=(H+L)\theta-HL,
\]
which gives $\E[\theta^2 | s]=(H+L)x-HL$. \qed

\subsection{Proof of Lemma \ref{lem:envelope}}


To establish this representation, compare the payoff formulas for two
consecutive message counts:
\[
W_n(x)-W_{n-1}(x)
=
\frac{2n-1}{3}
\left[
\frac{1}{4n^2(n-1)^2}-d^2x^2
\right].
\]
Recall that $t_n=\frac{1}{2dn(n-1)}$ is the threshold at which the equilibrium message count falls from $n$ to $n-1$.
Hence, $W_{ n - 1 } ( t_n ) - W_n ( t_n )$ is positive for $x < t_n$, zero for $x = t_n$, and negative for $x > t_n$.

Now fix $x\in[L,H]$ and let $m=N(x)$. By the definition of $N(x)$, $t_{m+1}\leq x<t_m$.
Because the thresholds $t_n$ decrease with $n$, for every $n\leq m$ we have
$x<t_n$. Therefore,
\[
W_m(x)>W_{m-1}(x)>\cdots>W_1(x).
\]
Likewise, for every $n\geq m+1$ we have $x\geq t_n$, and hence
\[
W_m(x)\geq W_{m+1}(x)\geq W_{m+2}(x)\geq\cdots.
\]
Thus, among all the quadratic payoff formulas, the largest value at $x$ is
attained by $W_m(x)=W_{N(x)}(x)$.
Since $N ( x ) \le N ( L )$ for all $x \in [ L , H ]$, it is therefore sufficient to consider the maximum in \eqref{eq:maxrepresentation}.

At $x=t_n$, the preceding payoff-difference formula gives $W_n(t_n)=W_{n-1}(t_n)$.
Therefore, $W$ is continuous when the equilibrium message count falls.
Immediately to the left of $t_n$, the selected payoff is $W_n$, whereas
immediately to the right it is $W_{n-1}$. Hence,
\[
    W'_{n-1}(t_n)-W'_n(t_n)
    =
    \frac{d(2n-1)}{3n(n-1)}>0.
\]

\qed

\subsection{Proof of Lemma \ref{lem:concavification}}
First, we show that two posterior means suffice.
Consider any feasible experiment inducing three or more
distinct posterior means with positive probabilities.
We can adjust the probabilities assigned to three of them
while preserving both their sum and their weighted mean.
Since the expected payoff is linear in these probabilities,
we can choose a direction of adjustment that weakly increases
the payoff and continue until one probability reaches zero.
Repeating this argument reduces the number of posterior means
to at most two without lowering the expected payoff.
Continuity of $W$ on $[L,H]$ ensures that the maximum
in \eqref{eq:concavification} is attained.

It remains to show that the optimal-value function equals
$\cav W$. The optimal-value function is concave because
mixing feasible distributions of posterior means at two
different means yields a feasible distribution at their
weighted average. It also lies weakly above $W$, since
no disclosure is feasible. Conversely, for any concave
function lying weakly above $W$, Jensen's inequality
implies that its value at $\mu$ is an upper bound on the
expected payoff of every feasible experiment.
Hence, the optimal-value function is the smallest
concave function lying weakly above $W$. \qed

\subsection{Proof of Proposition \ref{prop: sufficient condition for beneficial feedback}}


Consider $t_n \in ( L , H )$ as a prior mean and the following posterior spread $F_{ \varepsilon }$.
\begin{equation}
    X =
    \begin{cases}
    t_n - \varepsilon &\text{w.p. } 1/2 ,
    \\
    t_n + \varepsilon &\text{w.p. } 1/2 .
    \end{cases}
\end{equation}
For sufficiently small $\varepsilon > 0$, the value of $F_{ \varepsilon }$ is given by
\begin{align}
\begin{aligned}
    \Delta_W ( F_{ \varepsilon } , t_n )
    &= k_n \text{Var}_{ F_{\varepsilon} } ( X )
    + E_{ F_{\varepsilon} } [ W ( X ) - W_n ( X ) ]
    \\
    &= k_n \varepsilon^2
    +\frac{ 1 }{ 2 } \left[ W_{ n - 1 }' ( t_n ) \varepsilon + k_{ n - 1 } \varepsilon^2 - \left( W_n' ( t_n ) \varepsilon + k_n \varepsilon^2 \right) \right]
    \\
    &= \frac{ W_{ n - 1 }' ( t_n ) - W_n' ( t_n ) }{ 2 } \varepsilon + \frac{ k_{ n - 1 } + k_n }{ 2 } \varepsilon^2
    .
\end{aligned}
\end{align}
The first term is the difference of the marginal benefit of adjusting the message from $n$ to $n - 1$.
By Lemma 2, the difference $W_{ n - 1 }' ( t_n ) - W_n' ( t_n )$ is positive.
Then, $\Delta_W ( F_{ \varepsilon } , t_n ) > 0$ holds for sufficiently small $\varepsilon$; that is, the positive first-order gain from switching the number of messages dominates the second-order effect:
\begin{equation}
    \frac{ W_{ n - 1 }' ( t_n ) - W_n' ( t_n ) }{ 2 } \varepsilon + O ( \varepsilon^2 ) > 0 . 
\end{equation}
By continuity, this strict inequality holds in a neighborhood of $t_n$, corresponding to a nonempty open interval of priors. \qed

\subsection{Proof of Proposition \ref{prop:full}}

We first show that full disclosure is optimal for some interior prior if and only if $W^{FD}(x)\ge W(x)$
for all $x\in[L,H]$.

Suppose $W^{FD}(x)\ge W(x)$ for all $x\in[L,H]$.
Since $W^{FD}$ is affine, any feasible experiment satisfies
\[ \sum_j F_j W(x_j)
\le \sum_j F_j W^{FD}(x_j)
= W^{FD}(\mu), \]
where the equality follows from Bayes plausibility.
Since full disclosure achieves $W^{FD}(\mu)$, it is optimal for every interior prior.

Conversely, suppose $W(z)>W^{FD}(z)$ for some $z\in(L,H)$.
For any $\mu<z$, an experiment inducing posterior means $L$ and $z$ yields a payoff gain over full disclosure of
\[
\frac{\mu-L}{z-L}
[W(z)-W^{FD}(z)]>0.
\]
Similarly, for any $\mu>z$, an experiment inducing posterior means $z$ and $H$ yields a payoff gain of
\[
\frac{H-\mu}{H-z}
[W(z)-W^{FD}(z)]>0.
\]
At $\mu=z$, uninformative feedback strictly outperforms full disclosure.
Thus, full disclosure is strictly suboptimal for every interior prior.

Finally, the condition $W^{FD}(x)\ge W(x)$ for all $x\in[L,H]$ is equivalent to \eqref{eq:fullthreshold}, since equality holds at both endpoints and $(x-L)(H-x)>0$ for all interior $x$.

\subsection{Corollary \ref{prop:extremes}}

For statement 1, 
\begin{equation}\label{eq:fullconvexsufficient}
\alpha_R(2-\alpha_R)\ge\frac{d^2\{N(L)^2-1\}}3.
\end{equation}
 makes every quadratic $W_n$, $1\le n\le N(L)$, convex. Their pointwise maximum $W$ is then convex by Lemma~\ref{lem:envelope}. To see this directly, the value of each quadratic at a weighted-average belief is no larger than the weighted average of its endpoint values, which in turn is no larger than the corresponding weighted average of $W$. A convex $W$ lies below the line joining its endpoints, so Proposition~\ref{prop:full} establishes optimality.

It remains to show that this sufficient condition holds for small $d$. Feasibility at $L$ implies
\[
N(L)\{N(L)-1\}<\frac1{2dL},
\qquad
N(L)-1<\frac1{\sqrt{2dL}}.
\]
Since $N(L)^2-1=N(L)\{N(L)-1\}+N(L)-1$, we obtain
\[
0\le\frac{d^2\{N(L)^2-1\}}3
<\frac{d}{6L}+\frac{d^{3/2}}{3\sqrt{2L}}
\longrightarrow0\qquad\text{as }d\downarrow0.
\]
The left-hand side of \eqref{eq:fullconvexsufficient} is fixed and positive. Hence that condition holds for all sufficiently small $d>0$.

For statement 2, $dL\ge1/4$ rules out two or more positive-probability messages even at $x=L$. Thus $N(x)=1$ throughout $[L,H]$, and $W(x)=-1/12+\alpha_R(2-\alpha_R)x^2$ is strictly convex. \qed

 \subsection{Proof of Corollary \ref{prop: optimality of partial disclosure}}


Suppose first that there exists $n\geq2$ such that $t_n\in(L,H)$.
By Proposition~\ref{prop: sufficient condition for beneficial feedback}, informative feedback is beneficial for a nonempty open interval of priors.
Hence, for each prior in this interval, there exists a feedback experiment $F$ such that $\Delta_W(F,\mu)>0$.
Since condition~\eqref{eq:fullthreshold} fails, fully informative feedback is suboptimal for every interior prior.
Therefore, an optimal feedback experiment on this interval is neither uninformative nor fully informative, so partial disclosure is optimal.

Conversely, suppose that there is no $n\geq2$ such that $t_n\in(L,H)$.
Then $N(x)$ is constant on $[L,H]$, and the benchmark analysis applies.
Since fully informative feedback is strictly suboptimal, the benchmark characterization implies that uninformative feedback is optimal.
Hence, partial disclosure cannot be optimal for any prior. \qed

\subsection{Proof of Proposition \ref{thm:two}}\label{app:policyproof}


\paragraph{Step 1. Characterize the prior means at which uninformative feedback is optimal.}

For each $\mu\in[L,H]$, let $\mathcal F ( \mu )$ denote the set of all posterior-mean distributions that satisfy Bayes plausibility.
The distribution $\delta_\mu$ corresponds to uninformative feedback and yields $\Delta_W(\delta_\mu,\mu)=0$.
Define

\begin{equation}\label{eq:app-M0-delta}
\mathcal M_0
=\left\{\mu\in(L,H):
\Delta_W(F,\mu)\leq0\text{ for every }F\in\mathcal F(\mu)\right\}.
\end{equation}
By the definition of \(\mathcal M_0\) and the concavification representation, $x \in \mathcal M_0$ if and only if $V ( x ) = W ( x )$.
We refer to any $x$ satisfying $V(x)=W(x)$ as a \emph{contact point}.
Thus, $\mathcal M_0$ is precisely the contact set of $V$ and $W$.

First, consider an interior message-count threshold $t_n \in ( L , H )$.
By the proof of Proposition \ref{prop: sufficient condition for beneficial feedback}, for sufficiently small $\varepsilon>0$, we have $\Delta_W(F_\varepsilon,t_n) > 0$, where $F_{ \varepsilon } \in \mathcal F ( t_n )$ assigns equal probability to $t_n + \varepsilon$ and $t_n - \varepsilon$.
Therefore, $t_n \notin \mathcal M_0$, and all interior contact points lie within smooth message-count regions.

Next, let $\mu$ lie in a region with $n$ messages.
By the analysis of the benchmark, if $k_n > 0$, $\Delta_W ( F_{ \varepsilon } , \mu ) > 0$ for any sufficiently small $\varepsilon$.
Thus $\mu \notin \mathcal M_0$.
If $k_n=0$, $W_n$ is constant, say equal to $C$.
The upper-envelope representation in Lemma~\ref{lem:envelope}
implies $W(x)\geq C$ on $[L,H]$.
There must be some $z$ with $W(z)>C$: otherwise $W$ would
be globally constant and full disclosure would be optimal,
contrary to assumption.
Choose a point $y$ sufficiently close to $\mu$ on the side
opposite to $z$, within the same flat region.
A binary distribution supported on $\{y,z\}$ with mean $\mu$
then yields a strictly positive $\Delta_W$.
Consequently, if the prior $\mu$ in $\mathcal M_0$, then $k_{ N ( \mu ) } < 0$.

Finally, suppose that $a , b \in \mathcal M_0$ with $a < b$ satisfy $N ( a ) = N ( b ) = N$.
We show that every $\mu \in ( a , b )$ also belongs to $\mathcal M_0$.
By monotonicity of the equilibrium number of messages, we have $W = W_N$ throughout $[ a , b ]$.
Moreover, we have shown that $k_N < 0$, so $W_N$ is strictly concave.

Fix $\mu\in(a,b)$ and consider any binary experiment with posterior means $u < \mu < v$.
Define
\[
z (x)=\frac{v-x}{v-u}W(u)+\frac{x-u}{v-u}W(v) .
\]
Let $c=\max\{a,u\}$ and $d=\min\{b,v\}$.
Then, $c<\mu<d$, and $z (c)\leq W(c)$ and $z (d)\leq W(d)$.
Indeed, if $c=u$, the first inequality holds with equality; otherwise, $c=a$, and it follows from the optimality of uninformative feedback at $a$.
The second inequality follows analogously.
Concavity of $W_n$ on $[ a , b ]$ therefore gives
\begin{equation}
    W(\mu)
    \geq \frac{d-\mu}{d-c}W(c) + \frac{\mu-c}{d-c}W(d)
    \geq \frac{d-\mu}{d-c}z (c) + \frac{\mu-c}{d-c}z (d)
    =z (\mu).    
\end{equation}

Thus $\Delta_W(F_{u,v}^{\mu},\mu)\leq0$ for every such binary experiment. By the two-posterior representation, no experiment improves upon uninformative feedback at $\mu$, so $\mu\in\mathcal M_0$. Hence each message-count region intersects $\mathcal M_0$ in at most one interval, possibly empty or a singleton.

\paragraph{Step 2. Derive common-tangent conditions from the maximization of $\Delta_W$.}

Recall that $V ( \mu ) \equiv \sup_{F\in\mathcal F(\mu)} \E_F[W(X)]$ is the designer's optimized objective at prior $\mu$.
By Lemma~3, $V$ is the concave envelope of $W$.
Hence, uninformative feedback is optimal at $\mu$ if and only if $V ( \mu ) = W ( \mu )$.

Consider a maximal open interval $(a,b) \subseteq (L,H)$ on which $V > W$.
Since $V$ is the concave envelope of $W$, its endpoints are contact points with $W$, and $V$ is affine on $(a,b)$.
Therefore,
\begin{equation}
    V(x)
    =
    \frac{ b - x }{ b - a } W ( a ) + \frac{ x - a }{ b - a } W ( b ) ,
    \quad x\in[a,b].
\end{equation}
If both contact points lie in regions on which $W = W_n$ and $W = W_m$, the supporting line is tangent to $W$ at both endpoints:
\begin{equation}
    \label{eq:app-delta-tangency}
    W'(a)
    =
    W'(b)
    =
    \frac{W(b)-W(a)}{b-a}.
\end{equation}
If one endpoint is $L$ or $H$, the first-order condition applies only to the interior support point.

\paragraph{Step 3. Bound the number of disconnected uninformative-feedback regions.}

\begin{figure}[t]
\centering
\includegraphics[width=.8\linewidth]{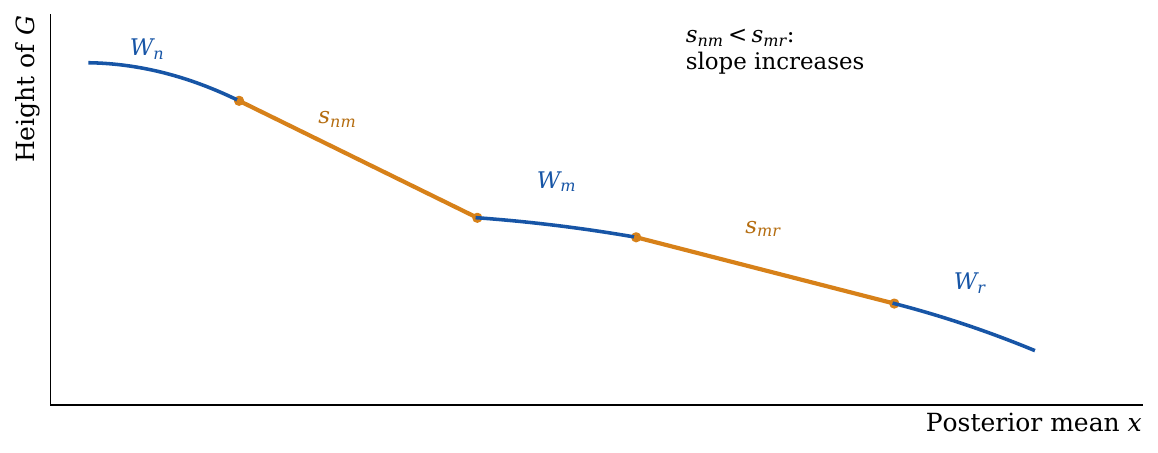}
\caption{The impossible three-contact configuration in Step~3 (schematic). The two orange affine gaps would have slopes $s_{nm}$ and $s_{ml}$. The common-tangent calculation forces $s_{nm} < s_{ml}$, an increase in slope from left to right, whereas a concave envelope must have non-increasing slopes. The blue arcs illustrate the three distinct contact regions.}\label{fig:impossible-contact}
\end{figure}

Suppose that two interior contact points $a < b$ lie in regions with the equilibrium number of messages $n$ and $m$, respectively.
Since $W_n(x)=-\frac{1}{12n^2}+k_nx^2$, the tangency conditions \eqref{eq:app-delta-tangency} imply
\begin{equation}
    2k_n a
    =
    2k_m b
    =
    \frac{W_m(b)-W_n(a)}{b-a}.
\end{equation}
Solving these conditions yields the slope of the common supporting line through the two contact points:
\begin{equation}
    s_{nm} = -\frac{\sqrt{k_nk_m}}{dnm}.
\end{equation}

Now consider three successive contact regions with message counts $n > m > l$.
The normalized quantity
\[
\frac{-k_j}{j^2}
=
\frac{d^2}{3}
-
\frac{\alpha_R(2-\alpha_R)+d^2/3}{j^2}
\]
is strictly increasing in $j$.
Hence the corresponding supporting slopes satisfy
\[
s_{nm}<s_{ml}.
\]
But the slopes of a concave function must be weakly decreasing from left to right.
Therefore, $V$ cannot have three distinct interior contact intervals.

\paragraph{Step 4. Establish a nondegenerate uninformative-feedback region and complete the classification.}

It remains to show that the contact set $\mathcal M_0$ contains a nondegenerate interval.
Suppose instead that every contact point were isolated.
Then, $V$ would be affine between any two consecutive contact points.
At an isolated interior contact point, concavity requires the left and right slopes of $V$ to coincide.
Hence the slopes on all maximal intervals on which $V$ is affine must coincide, so $V$ would be affine on $[L,H]$.

However, an affine concave envelope connecting the endpoints corresponds to full disclosure being optimal, contrary to the assumption under which full disclosure is suboptimal.
Therefore, $\mathcal M_0$ must contain at least one nondegenerate interval.

\paragraph{Step 5. Identify optimal posterior supports and derive the contact-point formulas.}

If a partial-disclosure region reaches the lower boundary $L$, the supporting line through $(L,W(L))$ is tangent to some quadratic piece $W_n$ at an interior contact point $a$.
Hence,
\begin{equation}
W_n'(a) = \frac{W_n(a)-W(L)}{a-L},
\end{equation}
which yields
\begin{equation}
a = L + \sqrt{ \frac{W(L)-W_n(L)}{-k_n} } .
\end{equation}
An analogous argument at the upper boundary gives
\begin{equation}
b = H - \sqrt{ \frac{W(H)-W_m(H)}{-k_m} }.    
\end{equation}
Thus, once the relevant contact regions are identified, the posterior supports of the optimal binary experiments are pinned down by tangency.
Thus it is globally optimal. This completes the proof of Proposition~\ref{thm:two}. \qed

\bibliography{bibtex} 

\end{spacing}
\end{document}